\documentclass[twocolumn,showpacs,prl,aps,amsmath,amssymb]{revtex4}

\usepackage{graphicx}
\usepackage{dcolumn}
\usepackage{bm}

\begin{document}
\title{Periodically rippled graphene: growth and spatially resolved
  electronic structure}

\author{A.L. V\'{a}zquez de Parga$^{1}$, F. Calleja$^{1}$, B. Borca$^{1}$,
  M.C.G. Passeggi Jr.$^{2}$, J.J. Hinarejos$^{1}$, F. Guinea$^{3}$ and R. Miranda$^{1}$}
\affiliation{$^{1}$Departamento de F\'{i}sica de la Materia
Condensada e Instituto de Ciencia de Materiales N. Cabrera.
Universidad Aut\'{o}noma de Madrid. Cantoblanco. 28049 Madrid.
Spain.} \affiliation{$^{2}$ Laboratorio de Superficies e
Interfases, INTEC (CONICET and UNL). S3000GLN Santa Fe.
Argentina.} \affiliation{$^{3}$ Instituto de Ciencia de Materials.
Consejo Superior de Investigaciones Cient\'{\i}ficas. Cantoblanco.
28049 Madrid. Spain.}

\date{\today}

\begin{abstract}
We studied the growth of an epitaxial graphene monolayer on
Ru(0001). The graphene monolayer covers uniformly the Ru substrate
over lateral distances larger than several microns reproducing the
structural defects of the Ru substrate. The graphene is rippled
with a periodicity dictated by the difference in lattice parameter
between C and Ru. The theoretical model predict inhomogeneities in
the electronic structure. This is confirmed by measurements in
real space by means of scanning tunnelling spectroscopy. We
observe electron pockets at the higher parts of the ripples.
\end{abstract}



\maketitle

The possibility to produce single layers of graphene
\cite{Novoselov,Geim} has opened a fascinating new world of
physical phenomena in two dimensions. Graphene has already shown
that its charge carriers are massless Dirac fermions
\cite{Novoselov1,Zhang,Novoselov2,Gusynin,Peres} and that it
displays an anomalous integer Quantum Hall Effect
\cite{Novoselov1,Zhang,Novoselov2} even at room temperature
\cite{Novoselov3}.
 Systems made up
of a few graphene layers can also be grown on a SiC substrate
\cite{Berger,Rutter}. Recently it has been shown that free
standing isolated graphene layer is intrinsically corrugated
\cite{Meyer}. These modulations in the heights of the graphene layers may be
related to the charge inhmogeneities observed in nominally undoped
samples\cite{Metal07} (see\cite{NG07}).

Ultra-thin epitaxial films of graphite have been grown on solid
surfaces for quite some time \cite{Oshima}. Even the growth of
"monolayer-graphite" films onto several substrates by Chemical
Vapor Deposition have been reported some time ago, but the degree
of characterization of the films was hampered by the existing
experimental limitations \cite{Hu}.

In this letter we report on a method to fabricate highly perfect,
on the scale of microns, periodically rippled graphene monolayer
and islands on Ru(0001) under Ultra High Vacuum (UHV) conditions.
We characterized by means of scanning tunnelling
microscopy/spectroscopy (STM/STS) the perfection at atomic scale
and the local electronic structure of the periodically rippled
graphene monolayer. The periodicity of the ripples is dictated by
the difference in lattice parameters of graphene and substrate,
and, thus, it is adjustable. We observe inhomogeneities in the
charge distribution, i.e electron pockets at the higher parts of
the ripples. This inhomogeneity in the electronic structure can be
understood with the help of a tight-binding model. The potential
associated with the rippled structure induce a charge transfer
from conduction to valence bands for some atoms and the opposite
in the others. Finally we studied graphene nanoislands that
display hexagonal shape with atomically resolved zigzag edges,
whose characteristic edge states allows us to determine the doping
of the graphene layer.

The experiments have been carried out in a UHV chamber with base
pressure of 4$\times$$10^{-11}$ Torr that contains a variable
temperature STM and a rear view Low Energy Electron Diffraction
(LEED) optics. The Ru(0001) crystal was cleaned by cycles of
$Ar^{+}$ sputtering and annealing followed to oxygen exposure and
heating to high temperature \cite{Calleja}. The graphene layers
were produced either by controlled segregation of C from the bulk
of the substrate or by thermal decomposition at 1000 K of ethylene
molecules pre-adsorbed at 300 K on the sample surface. The W tips
were routinely cleaned by ion bombardment and annealing. The
$dI/dV$ curves were obtained by numerical differentiation of the
$I(V)$ curves. We model the graphene electronic bands by a nearest
neighbor tight-binding model with parameter t=3eV, a periodic
shift of the carbon p$_z$ levels, and a finite broadening due to
the hybridization with the metallic substrate. In order to
calculate the density of states, we use a (30$\times$30) unit
cell, and a sum over 6 special points in the irreducible sector of
the Brillouin zone. Hence, the total number of states included in
the calculation is 10800. This is sufficient to resolve local
changes in the electron density in the order of
10$^{-2}$-10$^{-3}$ carriers per carbon atom.

The epitaxial layer of graphene covers completely the surface of
the single-crystal Ru substrate over distances larger than a
micron (see \cite{micron}) and presents a triangular periodicity
of 2.4 nm (Fig. 1(a)) that is due to the coincidence lattice of
graphene and Ru, i.e. the lattice of graphene is incommensurate
with the underlying lattice of Ru, with a size relation that
implies that 10 carbon honeycombs (0.246 nm) will adjust almost
exactly with 9 Ru-Ru interatomic distances (0.27 nm). The weakly
interacting (see below), laterally undistorted graphene structure
rides on top of the lattice of the substrate, resulting in some C
atoms being somewhat higher than others. The graphene layer, thus,
is rippled in a periodic fashion, with a periodicity dictated by
the difference in lattice parameters. Variations of this approach
using other experimental conditions can be used to produce
graphene monolayers on other single-crystal substrates
\cite{Fujita}.

With the STM images we characterize the type of defects that
appear in the epitaxial graphene layer. The monoatomic steps and
dislocations of the substrate are reproduced, but there are also
upper part of the ripples that are weaker or even missing
completely in the STM images, as can be seen in Fig. 1 (a) and in
ref. \cite{movie}. Fig. 1 (b) shows a high resolution topographic
STM image of the epitaxial graphene layer, revealing its atomic
structure. Unlike STM images of graphite \cite{Sinitsyna}, which
show normally only one of the two C atoms in the surface unit
cell, the honeycomb structure of graphene is clearly resolved in
the upper part of the ripples with its 0.14 nm C-C distance. The
Fourier transform of larger atomically resolved STM images shown
as inset in Fig. 1(b) indicates that the ripples constitute a
(10$\times$10) superlattice with respect to the C lattice. The
(9$\times$9) coincidence lattice with the Ru(0001) substrate is
revealed in the corresponding LEED pattern (not shown).

\begin{figure}
\includegraphics*[width=84mm]{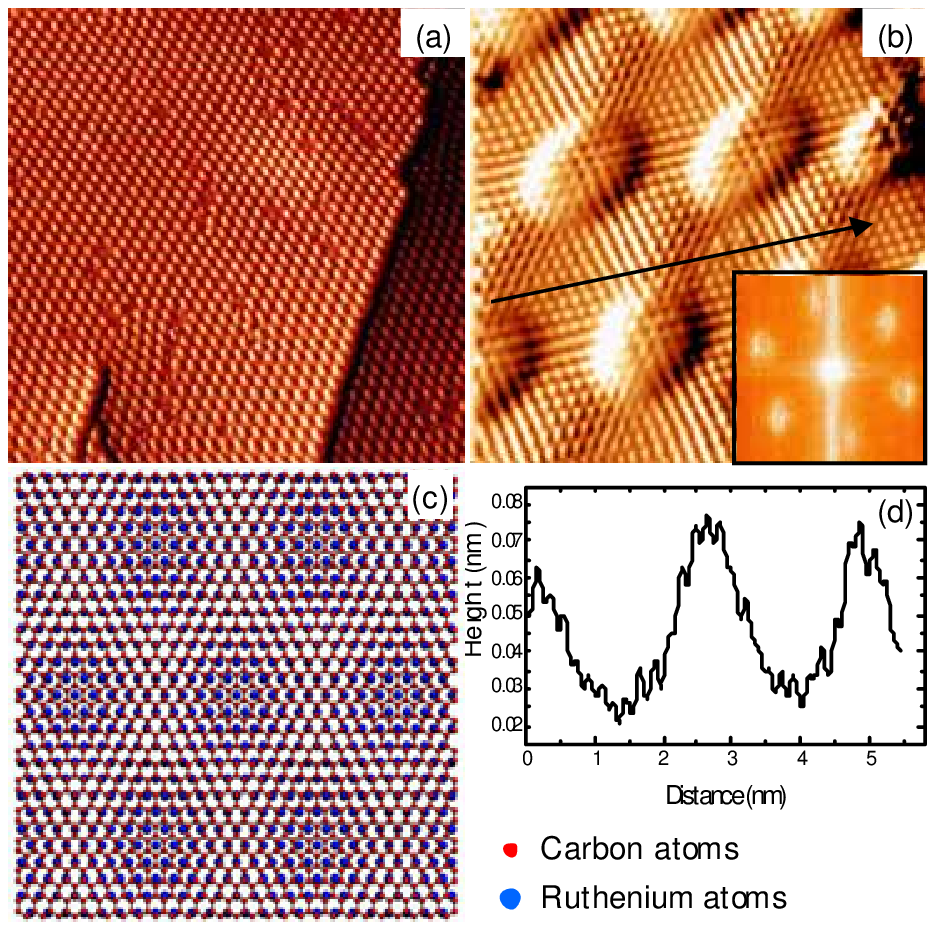}
\caption{(Color on line) : (a) 76 nm $\times$76 nm STM image of
graphene/Ru(0001) showing the decoration of a screw dislocation
and a monoatomic step from the substrate. There are also some
defects on the rippled structure. (b) 6.5 nm$\times$6.5 nm
atomically resolved image of graphene/Ru(0001). The image was
taken with a sample bias voltage of V$_{s}$=-4.5 mV and a tunnel
current of I$_{t}$=3 nA. The image is differentiated along the X
direction in order to see the weak atomic corrugation superimposed
to the ripples. The inset reproduces the Fourier transform of the
image showing the (10$\times$10) periodicity of the rippled
graphene layer. The larger hexagonal pattern corresponds to the
C-C distances and the smaller spots to the periodic ripples. (c)
Corresponding structural model. (d) Line profile marked with an
arrow in panel (b). The atomic corrugation is around 5 pm.}
\label{figure1}
\end{figure}

The apparent vertical corrugation of the rippled graphene
monolayer, as seen with STM, changes with the tunnelling voltage,
ranging from 0.1 nm (at a sample voltage of -1 V) to 0.02 nm (at
+1 V). This indicates the importance of electronic effects. The
C-C apparent atomic corrugation inside the honeycomb unit cell is
only of the order of 0.005 nm (Fig 1(d)). The actual geometric
corrugation of the rippled layer can be estimated to be below 0.02
nm from the observed lack of diffracted peaks for a beam of
thermal He atoms scattering off the graphene surface \cite{Nieto}.

The bonding with the substrate occurs through the hybridization of
the C $\pi$-states with the Ru \textit{d} states. Photoelectron
spectroscopy shows clearly that the layer bonding is not carbidic.
Angular Resolved Photo Electron Spectroscopy data show that the
graphene bands in graphene/Ru(0001) are similar to the ones of
graphite, but rigidly shifted down in energy \cite{Himpsel}. The
bottom of the $\pi$ band at the center of the Brillouin Zone is
shifted down by 1.8 eV with respect to graphite \cite{Himpsel}.
The small energy shift of the C\textit{1s} core level with respect
to graphite indicates that the charge transfer from the substrate
is small, but not negligible, i.e. the graphene layer is doped
with electrons by the substrate.

The graphene-Ru interface is atomically well defined. The average
interlayer distance for the closely related incommensurate
graphene/Pt(111) system has been determined to be rather large
(0.37 nm) by means of tensor LEED analysis \cite{Hu}. First
principles calculations in the similar graphene/Ir(111) system
have indicated that the average binding (0.2 eV/C atom) is barely
strong enough to correspond to chemical bond formation
\cite{Diaye}. The incommensurate nature of the graphene monolayer
is an additional indication that the interaction of graphene with
the substrate is rather weak.

Fig. 2(a)  shows spatially-resolved $dI/dV$ tunnelling spectra,
which are roughly proportional to the Local Density of States
(LDOS),  recorded on top of the "high" and "low" regions of the
corrugated graphene layer. Panel (b) shows the corresponding
calculation. The experimental tunnelling spectra recorded at
different spatial positions are obviously different: the occupied
LDOS is systematically larger in the "high" areas of the rippled
layer, while the empty LDOS is larger in the "low" parts. The
differences are robust enough to survive at 300 K. There are also
weaker features at both sides of the Fermi level, separated from
each other by 0.3 eV and located at about the same energies in
both regions of the ripples.

\begin{figure}
\includegraphics*[width=84mm]{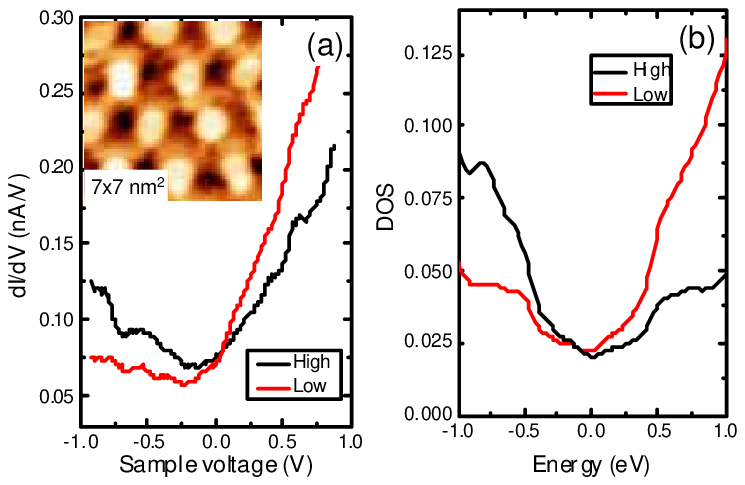}
\caption{(Color on line) (a) $dI/dV$ curves recorded at 300 K on
the higher (black curve) and lower (red curve) areas of the
rippled graphene monolayer. The inset shows the topographic image
measured simultaneously. (b) Calculations of the spatially
resolved LDOS for an isolated, (10$\times$10) periodically
corrugated graphene layer. The spectra correspond to the high
(black curve) and low (red curve) regions of the rippled graphene
layer.} \label{figure2}
\end{figure}

The electronic structure has been simulated by taking into account
that the rippled graphene layer is not too distorted. The model
calculations have been performed for an isolated graphene layer in
which the effect of the substrate has been considered to result
in: i) a shift of the Dirac point by -0.3 eV due to doping; ii)
the introduction of a finite lifetime caused by hybrization of the
$\pi$ orbitals with the band of the substrate; and iii) a
(10$\times$10) periodic potential that changes between -3V and
3V/2,where V=-0.3eV to account for the periodic structural
ripples. This value for the potential gives the better agreement
with the experiments. For higher values of the potential the
asymmetry in the calculated DOS is bigger than the one found in
the experiments. The $\pi$ band in graphene has a total width of
W$\sim$6t, where t$\sim$3 eV is the hopping between $\pi$ orbitals
at nearest neighbor C atoms \cite{Brandt}.

In agreement with the experiments, the calculations show that the
occupied LDOS is larger on the "high" regions of the superlattice,
where the potential is at a minimum, while the empty LDOS is
larger at the "low" regions of the graphene layer(see lower panel
in Fig. 3). This effect is extremely robust and indicates that the
valence band is depleted in the low portions of the ripples, while
the conduction band is depleted at the high parts of the ondulated
graphene layer. The calculations also reproduce the weaker
features at both sides of the Fermi energy that reflect the
existence of the periodic superlattice, which induces a folding of
the graphene bands in the new (10$\times$10) Brillouin Zone. The
reconstruction consider here does not superimpose the two
inequivalent corners of the Brillouin zone and it does not open a
gap at the Dirac energy \cite{Manes}. The spectrum is split on
subbands separated by gaps, away from the K and K' points.

The periodic charge inhomogeneities in the graphene layer can be
visualized directly in the real space by imaging the spatial
distribution of $dI/dV$ close to the Fermi energy. Fig. 3 shows
the spatial distribution of the LDOS below and above the common
Fermi level (which corresponds to the shifted neutrality level of
the doped graphene layer). The experimental images are in the
upper row at the left and the right of the corresponding
topographic image. The bright regions correspond to larger LDOS.
For a complete set of spatially resolved $dI/dV$ maps versus
sample bias voltage see ref. \cite{movie}.

\begin{figure}
\includegraphics*[width=84mm]{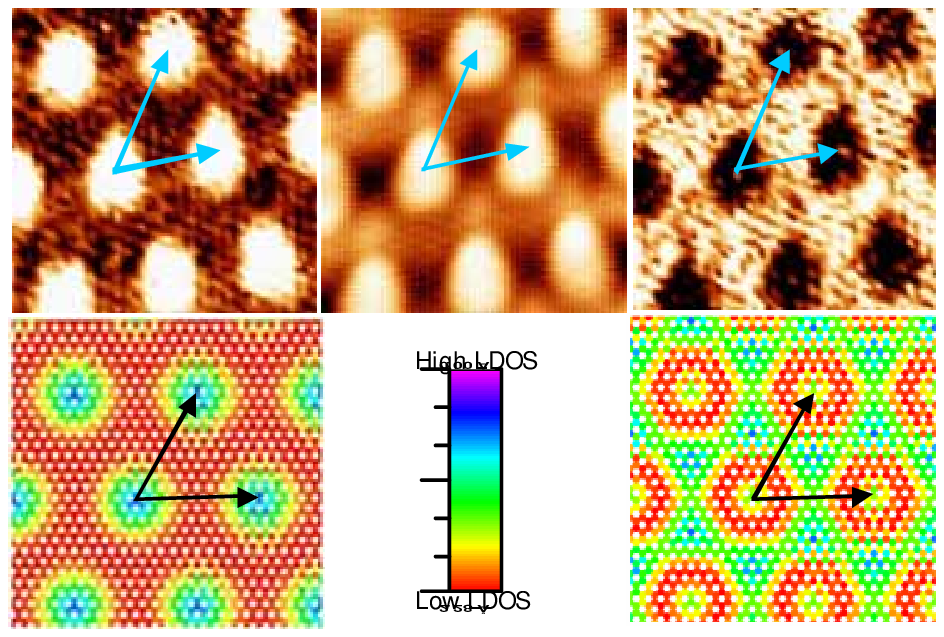}
\caption{(Color on line) The left and right images in the upper
panel are maps of $dI/dV$ at -100 meV and + 200 meV and reflects
the spatial distribution of the LDOS  below and above the Fermi
level, respectively,  for an extended graphene layer on Ru(0001).
The central image shows the topographic image recorded
simultaneously. The lower panel shows the corresponding
calculations for the spatially resolved LDOS for a (10$\times$10)
periodically corrugated graphene layer. The arrows indicate the
(10$\times$10) unit cell.} \label{figure3}
\end{figure}

These inhomogeneities in the charge distribution produced by the
periodic ripples are probably also present in the structural
corrugations inherent to free standing, isolated graphene
monolayers \cite{Meyer}.

The growth method described here can also be used to produce
different graphene nanostructures, such as nanowires on vicinal
surfaces or nanometer wide islands as illustrated in Fig. 4(a).
The apparent step height of the island is 0.15 nm (from the Ru
surface to the lower part of the ripples) showing that the
graphene layer is indeed only one monolayer high. The inset shows
that the islands are truncated hexagons with straight edges of a
single structural type. It is remarkable that the periodically
rippled coincidence lattice goes right to the steps of the
islands. In fact in some cases the island step cuts the
(10$\times$10) arrangement. This confirms that the overlayer of
graphene is indeed weakly coupled to the Ru substrate. Further
zooming into the  island reveals that the geometry of the steps is
of the zigzag type.

\begin{figure}
\includegraphics*[width=84mm]{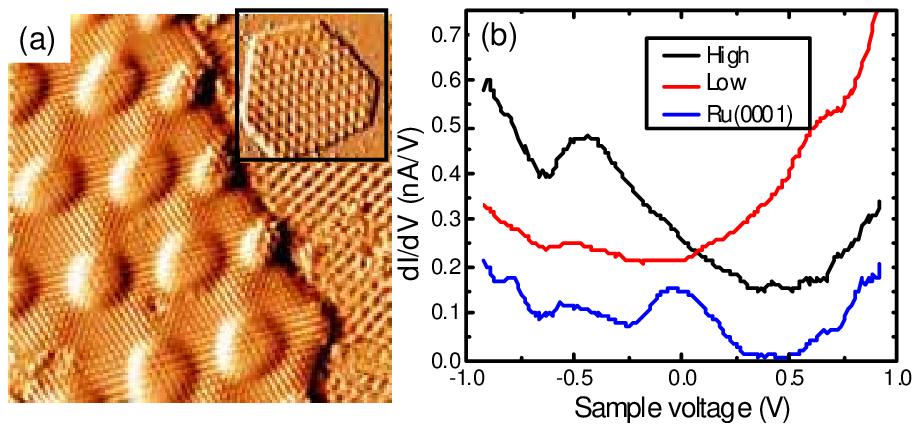}
\caption{ (Color on line)(a) 10 nm$\times$10 nm atomically
resolved STM image of a graphene island on Ru(0001). The image was
recorded at V$_{s}$=-4.5 meV and I$_{t}$=3 nA. The inset shows an
image of the whole island with a lateral size of 47 nm. The images
are differentiated in X direction. (b) Spatially resolved
tunnelling spectra measured on the high (black curve) and low (red
curve) areas of the ripples close to the edge of the island and on
clean Ruthenium (blue curve). The Ruthenium spectra was offset for
clarity} \label{figure4}
\end{figure}

The relative orientation of the graphene layer with respect to the
underlying Ru lattice can be determined by resolving
simultaneously both atomic lattices. In order to facilitate the
resolution of the weakly corrugated Ru lattice, CO has been
adsorbed at saturation in the parts of the Ru surface not covered
by graphene. The geometry of the resulting
($\sqrt{3}\times\sqrt{3}$ R30$^{\circ}$) structure of CO on
Ru(0001) is well known \cite{Gsell}, and, thus, the relative
orientation of graphene and Ru lattices represented in Fig. 1 has
been obtained.

Fig. 4(b) shows local tunnelling spectra recorded when the surface
is only partially covered with graphene. The advantage of this
situation is that one can take spectra in the clean Ru patches of
the surface, where only the surface state of Ru should be
detected, and, thus, spurious features from the tip DOS can easily
be detected. The spectra recorded at the center of the islands are
undistinguishable from the one obtained on extended graphene
layers (see Fig. 2). The spectra taken at the edges, however, show
a new peak at -0.4 eV, particularly on the "high" areas of the
ripples, which might be related to the theoretically predicted
edge states \cite{edge}. The spectrum recorded on the clean Ru
patches show only the surface state around the Fermi energy
observed on Ru(0001) \cite{Calleja}.

In summary we have grown graphene monolayers and islands on
Ru(0001). The graphene monolayer presents a periodically rippled
surface due to the difference in lattice parameter between
graphene and the metallic substrate. The periodic ripples produce
a spatial charge redistribution in the graphene. This have been
measured with spatially resolved $dI/dV$ maps and confirmed with a
theoretical model. The new periodicity also induce the opening of
a series of minigaps by the additional periodic potential. The
presence of these minigaps is expected to give rise to new
phenomena at low temperatures in the presence of high magnetic
fields. We atomically resolve the zig-zag edges of graphene
islands and we use the position in energy of the edge state to
determine the doping of the graphene layer. The methods described
here can be implemented on many other single crystal substrates,
giving rise to a series of graphene monolayers with different,
substrate-dependent, periodic corrugations and, thus, opening the
possibility to systematically test the electronic properties of
controlled, charge inhomogeneous graphene layers. Methods to
transfer the present finding to experimental systems adequate for
lateral transport measurements are in progress.

Partial financial support by the Ministerio de Educaci\'{o}n y
Ciencia through projects MAT2003-08627-C02-02,
NAN2004-08881-C02-01, FIS2005-05478-C02-01, the Comunidad de
Madrid, through the programs CITECNOMIK, CM2006-S-0505-ESP-0337
and NANOMAGNET, S-0505/MAT/0194 and the European Union Contract
12881 (NEST),  is gratefully acknowledged.



\begin{thebibliography}{10}

\bibitem{Novoselov} K.S. Novoselov et al., Science \textbf{306}, 666 (2004).

\bibitem{Geim} A.K. Geim and K.S. Novoselov, Nature Mater.
\textbf{6}, 183 (2007).

\bibitem{Novoselov1} K.S. Novoselov et al., Nature \textbf{438},
197 (2005).

\bibitem{Zhang} Y. Zhang et al., Nature \textbf{438}, 201 (2005).

\bibitem{Novoselov2} K.S. Novoselov et al., Nature Phys.
\textbf{2}, 177 (2006).

\bibitem{Gusynin} V.P. Gusynin and S.G. Sharapov, Phys. Rev.
Lett. \textbf{95}, 146801 (2005).

\bibitem{Peres} N.M.R. Peres, F. Guinea and A.H. Castro Neto,
Phys. Rev. B \textbf{73}, 125411 (2006).

\bibitem{Novoselov3} K.S. Novoselov et al., Science \textbf{315},
1379 (2007).

\bibitem{Berger} C. Berger et al., J. Phys. Chem. B \textbf{108},
19912 (2004).

\bibitem{Rutter} G.M. Rutter et al., Science \textbf{317}, 219
(2007).

\bibitem{Meyer} J.C. Meyer et al., Nature \textbf{446}, 60 (2007).

\bibitem{Metal07} J. Martin et al., arXiv:0705.2180.

\bibitem{NG07} A. H. Castro Neto and E. A. Kim, arXiv:cond-mat/0702562, F. Guinea,
M. I. Katsnelson, and M. A. H. Vozmediano, arXiv:0707.0682.

\bibitem{Oshima} For a review see Ch. Oshima and A. Nagashima, J.
Phys.: Condens. Matter. \textbf{9}, 1 (1997).

\bibitem{Hu} Z.-P. Hu et al., Surf. Sci. \textbf{180}, 433 (1987).

\bibitem{Calleja} F. Calleja et al., Phys. Rev. Lett. \textbf{92}
206101 (2004).

\bibitem{micron} See EPAPS Document No. [] for a STM 1$\mu$m $\times$ 1$\mu$m image of graphene on Ru(0001).

\bibitem{Fujita} T. Fujita, W. Kobayashi and C. Oshima, Surf.
Interf. Analysis, \textbf{37}, 120 (2005).

\bibitem{Sinitsyna} O.V. Sinitsyna and I.V. Yaminsky, Russian,
Chem. Rev. \textbf{75}, 22 (2006).

\bibitem{Nieto} P. Nieto et al. (in preparation).

\bibitem{Himpsel} F.J. Himpsel et al., Surf. Sci. Lett.
\textbf{115}, L159 (1982).

\bibitem{Diaye} A.T. N'Diaye et al., Phys. Rev. Lett. \textbf{97},
215501 (2006).

\bibitem{Brandt} N.B. Brandt, S.M. Chudinov and Y.G. Ponomarev, in
Modern Problems in Condensed Matter Sciences, ed. V.M. Agranovich
and A.A. Maradudin (North Holland 1988) Vol. 29.1

\bibitem{Manes} J.L. Ma\~nes, F. Guinea and M.A.H. Vozmediano, Phys.
Rev. B \textbf{75}, 155424 (2007).

\bibitem{movie} See EPAPS Document No. [] for a complete
set of spatial resoved $dI/dV$ maps versus sample bias voltage in
an energy range of $\pm$1 V.

\bibitem{Gsell} M. Gsell, P. Jakob and D. Menzel, Science
\textbf{280}, 717 (1998).

\bibitem{edge} Note that deviations from electron-hole symmetry away from the
Dirac energy shift these states away from the neutrality point,
see N.M.R. Peres et al. \cite{Peres}.

\end{thebibliography}
\end{document}